\documentclass[prl,superscriptaddress,showpacs,twocolumn]{revtex4}

\usepackage{graphicx}

\usepackage{graphicx,amsfonts}

\usepackage{epsfig,amsmath}

\usepackage{verbatim}

\begin{document}

\newcommand{\atanh}
{\operatorname{atanh}}

\newcommand{\ArcTan}
{\operatorname{ArcTan}}

\newcommand{\ArcCoth}
{\operatorname{ArcCoth}}

\newcommand{\Erf}
{\operatorname{Erf}}

\newcommand{\Erfi}
{\operatorname{Erfi}}

\newcommand{\Ei}
{\operatorname{Ei}}

\newcommand{\sgn}{{\mathrm{sgn}}}
\newcommand{\rme}{{\mathrm{e}}}
\newcommand{\rmd}{{\mathrm{d}}}
\def\be{\begin{equation}}
\def\ee{\end{equation}}

\def\bea{\begin{eqnarray}}
\def\eea{\end{eqnarray}}

\def\e{\epsilon}
\def\l{\lambda}
\def\d{\delta}
\def\o{\omega}
\def\cb{\bar{c}}
\def\Li{{\rm Li}}

\title{The universal high temperature regime of pinned elastic objects }

\author{Sebastian Bustingorry}
\affiliation{CONICET, Centro At{\'{o}}mico Bariloche, 8400 San Carlos de
Bariloche, R\'{\i}o Negro, Argentina}
\author{Pierre Le Doussal}
\affiliation{CNRS-Laboratoire de Physique
Th{\'e}orique de l'Ecole Normale Sup{\'e}rieure, 24 rue Lhomond, 75231
Paris Cedex, France}
\author{Alberto Rosso}
\affiliation{CNRS-Universit\'e Paris-Sud, LPTMS, UMR8626-B\^at 100,91405 Orsay Cedex, France}

\date{\today}

\begin{abstract}
We study the high temperature regime within the glass phase of an elastic object with short ranged disorder. In that regime we argue that the scaling functions of any observable are described by a continuum model with a $\delta$-correlated disorder and that they are universal up to only two parameters that can be explicitly computed. This is shown numerically on the roughness  of directed polymer models and by dimensional and functional renormalization group arguments. We discuss experimental consequences such as non-monotonous behaviour with temperature. 
\end{abstract}

\maketitle

Elastic objects pinned by quenched disorder are ubiquitous in nature, e.g. vortex lattices in superconductors \cite{vortex}, magnetic domain walls \cite{lemerle}. They are modeled by elastic manifolds, of internal dimension $d$, parameterized by a $N$-component displacement field $u(x)$, submitted to random potentials. These systems have been described using the collective pinning theory \cite{LO}, in terms of the characteristic Larkin length $R_c$, and more recently using functional renormalization group (FRG) \cite{BalentsFisher,frg} in terms of a RG fixed point at zero temperature, with only few universality classes depending on $N,d$ and the nature of elasticity and disorder - short range (SR) or long range (LR). For scales larger than $R_c$ these objects exhibit glass phases with statistically self-similar ground states of roughness scaling as $u \sim x^\zeta$ and self-similar energy landscape with free energy exponent $\theta$. Whenever $\theta>0$ the $T=0$ fixed point is attractive and thermal fluctuations are irrelevant at low temperature for large systems. In some cases ($\theta_F>0$ see below) the glass phase extends to all temperatures, with, however, a crossover at $T=T_{dep}$. At high temperature $T>T_{dep}$,  the system unbinds from individual pins but remains collectively pinned, and the Larkin length increases with $T$ \cite{nattermannshapir,vortex,mueller2001}. This has interesting consequences, e.g. a reentrant region in the phase diagram of high $T_c$ superconductors \cite{ertas}. It was generally argued that, due to this effect, the roughness amplitude decays with temperature for SR disorder \cite{nattermannshapir,us3,agoritsas}. 

Besides its interest to experiments, the high temperature regime is also at the center of recent works on the directed polymer (DP) $d=1$, a problem in close connection to KPZ growth and Burgers turbulence \cite{directedpoly,spohnreview,spohnnewKPZ,Burgers}. In two dimension ($N=1$) it is amenable to the Bethe Ansatz method using a continuum model with $\delta$-function correlation in the random potential \cite{BAold}. Since the collective pinning theory and the FRG show that a finite correlation range in the disorder is an important ingredient to describe pinning, an outstanding question has been the domain of validity of this model. Recently the distribution of the free energy $F$ of polymers of length $x$ was computed within the $\delta$-correlated model \cite{us3,dotsenko,spohnnewKPZ,math}. At large scale $x$ one recovers the Tracy Widom distribution \cite{TW1994}, previously proved to hold \cite{Johansson2000} for a discrete DP model, but at $T=0$. At first sight, it suggests that the $\delta$-function model also captures the universality {\it at low temperature}, in agreement with the idea of an effective scale dependent temperature flowing to zero at large scale. However, as emphasized in \cite{us3}, the universal results obtained from the $\delta$-function model are a priori correct only in the infinite temperature limit $T \to + \infty$ with $\tilde u=u/T^3$ and $\tilde x=x/T^5$ being kept fixed. Although the scaled distribution of $F$ in the large size limit does not appear to depend on $T$, its variance $\overline{F^2}^c \sim A(T) x^{2 \theta}$ exhibits an {\it amplitude} $A(T)$ which could be computed only at high $T$ and it is found to decay with temperature \cite{us3}. 

In this Letter we reexamine the universal high temperature regime for the DP and for general pinned elastic objects. We first argue from dimensional analysis that, in all cases where the disorder is SR and the glass phase extends to all temperatures, the large $T$ limit is described by a continuum model with gaussian and $\delta$-correlated disorder. It contains only two parameters $\kappa$, the elasticity, and $g$ the amplitude of the disorder. We then demonstrate numerically that at high temperature the $x$-dependent roughness of the DP for various discrete models, falls onto a universal curve when expressed in the rescaled variables. The fit involves no other adjustable parameters than $g$ and $\kappa$ which, furthermore, can be {\it computed explicitly} for each discrete model, leading to a very predictive theory. Next, we show within the FRG how to account for this high $T$ universality, and we recover the temperature dependence of the rescaled variables which, as anticipated in \cite{nattermannshapir} involves the Flory exponents. 

We consider the standard model for pinning of partition function $Z = \int {\cal D}[u] e^{- H/T}$ with:
\begin{equation} \label{standard}
 H =  \int_x \left[ \frac{\kappa}{2} \left(\nabla_{x} u(x) \right)^2 + V(u(x),x) \right],
\end{equation}
where $\int_x=\int d^d x $. The second cumulant of the random potential is given by $\overline{V(u,x) V(u',x')}^c = g \delta^d(x-x') R(u-u')$, where $R(u)$ is a short range function with $\int d^N u R(u)=1$. A small scale cutoff $x_0$ is implicit whenever needed. For zero average gaussian random potential one finds  using replica that $\overline{Z^p} = \int {\cal D}[u] e^{- S^{rep}}$ with 
\bea \label{srep}
 S^{rep} =    \frac{\kappa}{2 T}  \sum_{\alpha=1}^p \int_x \left(\nabla_x u_\alpha(x) \right)^2  
 - \frac{g}{2T^2}  \sum_{\alpha \beta=1}^p \int_x  R(u_{\alpha,\beta}),
\eea
where $u_{\alpha,\beta}=u_\alpha(x)- u_\beta(x)$. To study the high $T$ regime we first substitute the rescaled coordinates and fields, $x=b \tilde{x}$,   $u= a \tilde u$ in the replicated action (\ref{srep}).  For $d<2$ the small scale cutoff $x_0$ can be set equal to zero and the temperature dependence is removed by the choice:
\begin{equation} \label{conditions}
 \frac{\kappa b^{d-2} a^2}{T} = 1 \qquad \frac{b^d g}{T^2 a^N}=1,
\end{equation}
which is solved as:
\begin{equation} 
 a= \left(\frac{\kappa^2}{g}\right)^\frac{\theta_F - 2 \zeta_F}{\theta_F(4+N)} \left(\frac{T}{\kappa}\right)^{\frac{\zeta_F}{\theta_F}}, 
  \,  b =  \left(\frac{\kappa^2}{g}\right)^\frac{2}{\theta_F(4+N)}  \left(\frac{T}{\kappa}\right)^{\frac{1}{\theta_F}},
\label{solution}
\end{equation}
where the first condition in (\ref{conditions}) requires $\theta_F = d-2 + 2 \zeta_F$, and the second $\zeta_F=(4-d)/(4+N)$, i.e. the Flory roughness and free energy exponents.  For  $d\ge 2$, as discussed later,  the presence of a small, but finite cut-off $x_0$ slightly modifies Eq.(\ref{conditions}) and Eq.(\ref{solution}) remains valid if we replace $\theta_F$ by $2 \zeta_F $ leading to $a=(T/\kappa)^{1/2}$ and $b= R_g (T/\kappa)^{1/(2 \zeta_F)}$ where $R_g=(\frac{\kappa^2}{g})^\frac{1}{4-d}$ is similar to the $T=0$ Larkin length. In the glass phase, where $\theta_F,\zeta_F>0$, the high temperature regime corresponds to large $a$.  Using $ \lim_{a \to \infty} a^N R(a \tilde u) =  \delta^N(\tilde u)$, the model becomes equivalent to:
\bea \label{delta}
\tilde S^{rep} = \int_{\tilde x} \frac{1}{2} \sum_{\alpha=1}^p \left(\nabla_{\tilde{x}} \tilde u_\alpha(\tilde{x}) \right)^2 
 - \frac{1}{2} \sum_{\alpha \beta=1}^p \int_{\tilde x} \delta^N(\tilde u_{\alpha,\beta}),
\eea
i.e. a continuum model where $\kappa,T$ and $g$ have been set to unity and the disorder is $\delta$-correlated. Note that at the thermal fixed point $T=\infty$, the disorder term scales as $x^{\frac{1}{2} \theta_F(4+N)}=x^{\frac{2 d + N(2-d)}{2}}$. If $\theta_F>0$, the glass phase exists at all temperature, the case considered here.  

Let us consider the high temperature regime for much more general models. First one can break the statistical tilt symmetry (STS) $u_\alpha(x) \to u_\alpha(x)+\phi(x)$, which implies the non-renormalization of $\kappa$, by considering e.g.  $\overline{V(u,x) V(u',x')}^c = R(u-u',x-x')$. Then one easily sees that the same rescaling yields $\lim_{a \to \infty, b \to \infty} a^N b^d R(a \tilde u,b \tilde x) =  \delta^d(\tilde x) \delta^N(\tilde u)$.  Similarly a third cumulant of disorder yields a $g_3 \delta(\tilde u_{\alpha,\beta}) \delta(\tilde u_{\alpha,\gamma})$ interaction in the replicated action but its strength decays as $g_3/g \sim 1/(T a^N)$ at high $T$, and similarly for higher cumulants. It is thus quite reasonable to expect that provided a model is SR it will fall at high $T$ in the same universal behavior as the model (\ref{delta}) (corresponding to  gaussian disorder) hence will be also described  by universal scaling functions up to 2 parameters $\kappa$ and $g$. 

\begin{figure}[!tbp]
\includegraphics[width=3.8cm,clip=true]{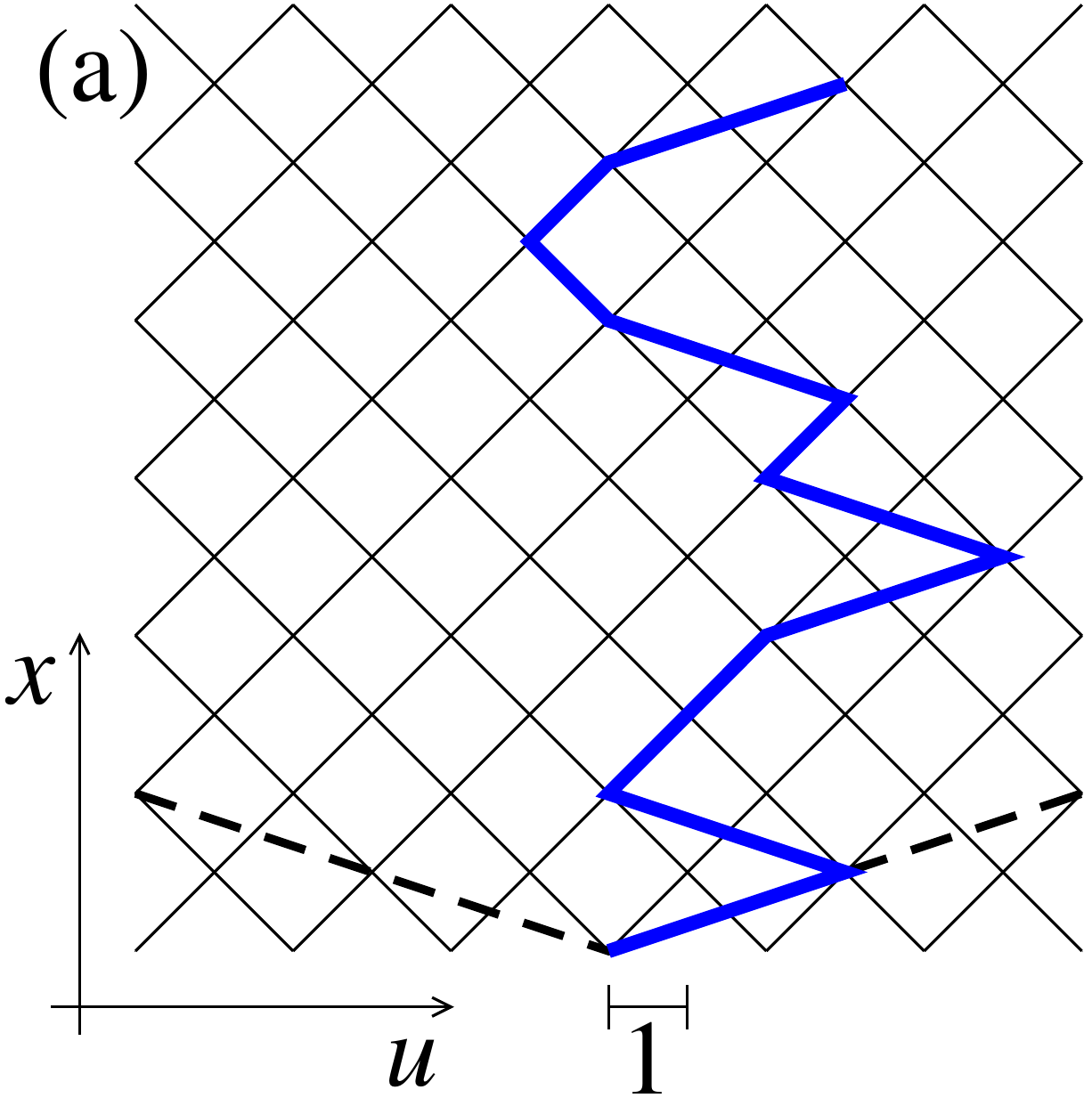}
\includegraphics[width=4.2cm,clip=true]{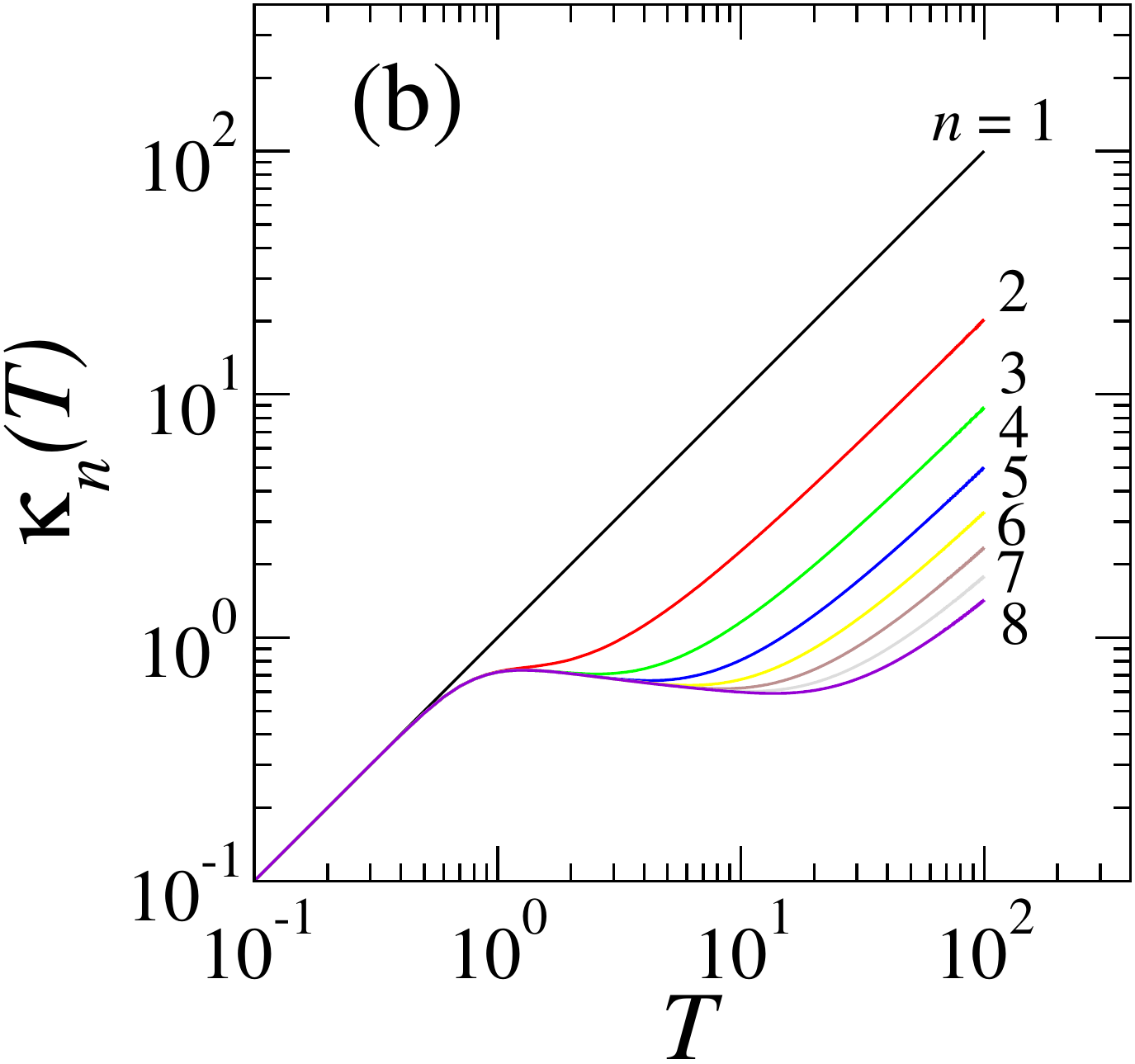}
\caption{\label{fig:model} (Color online) (a) Sketch of the model. Solid line corresponds to a polymer with elongations up to $n=2$. (b) Temperature dependence of $\kappa_n$ for different $n$.}
\end{figure}

We now explore the consequences of Eq. (\ref{delta})  on the DP ($d=N=1$) using numerics. We recall that, for this model, $\zeta=2/3$, $\theta=1/3$ and the Flory exponents are $\zeta_F=3/5$ and $\theta_F=1/5$. For clarity, we focus on the behavior of the roughness in the the so-called droplet geometry where a line of size $x$ has one end pinned at the origin ($u(0)=0$) and the other one is free. Similar conclusions  are expected also for other boundary conditions and for other observables, like the fluctuation of the free energy. A directed polymer is an elastic line living on a square lattice with a solid-on-solid restriction $|u(x+1)-u(x)| = 1$. Impurities of  energy $V(u,x)$ are drawn on each site of the lattice. The sum of the energies associated to the sites linked by a polymer defines the energy of the polymer. Here we release the solid-on-solid restriction allowing longer elongations $|u(x+1)-u(x)| = 2 j-1$ ($j=1,\ldots,n$) with an elastic cost of $j^2$. Figure~\ref{fig:model}(a) shows the geometry of the model for $n=2$. For $n=1$ we recover the standard directed polymer.  These lattice models violate STS because the restriction on the local elongation generates  non-harmonic terms in the elastic energy. For this reason, at low temperature, we expect a renormalization of both $\kappa$ and $g$ taking a different value respect to the bare ones. A major simplification occur in the high temperature regime where one can explicitly compute $\kappa$ and $g$. In practice, $\kappa$ can be extracted from the model without disorder for which the polymer  behaves like a particle diffusing on a one dimensional lattice ($u(x)$ being the particle "position" at  "time" $x$).  The mean square displacement of the particle  is given by $\langle u^2(x) \rangle_{T,n}=T x / \kappa_n $. The ratio $T / \kappa_n$ coincides with mean square jump of the particle,  i. e. with the mean square local elongation of the polymer, hence:
\begin{equation}
 \kappa_n(T) = \frac{\sum_{j=1}^n e^{-\beta j^2}}{\sum_{j=1}^n \beta (2j-1)^2 e^{-\beta j^2}}.
\end{equation}
The temperature behavior of $\kappa_n$ is shown in Fig.~\ref{fig:model}(b). At low temperatures, only the smallest elongation is actived and $\kappa_n(T \to 0) =\kappa_1=T$. At high $T$ all allowed elongations are equally activated and $\kappa_n (T \to \infty)= 3T/(4 n^2-1)$. The continuum elastic string limit is achieved for large $n$ and $T$ with $n^2 \gg T$, and $ \kappa_n \to \kappa = 1/2$. The parameter $g= \sum_{(u,x),(u',x')}\overline{V(u,x)V(u',x')}^c$. Two impurity distributions were implemented:  i)  $V(u,x)$ are uncorrelated Gaussian numbers of variance $g$ ii) "Poissonian" disorder where each site contains zero or one impurity (with probability $p$) of fixed strength $d$. In the latter case although all the disorder cumulants are non zero, only the second one, with $g=d^2 p (1-p) $, remains relevant in the large temperature limit.

\begin{figure}[!tbp]
\includegraphics[width=8cm,clip=true]{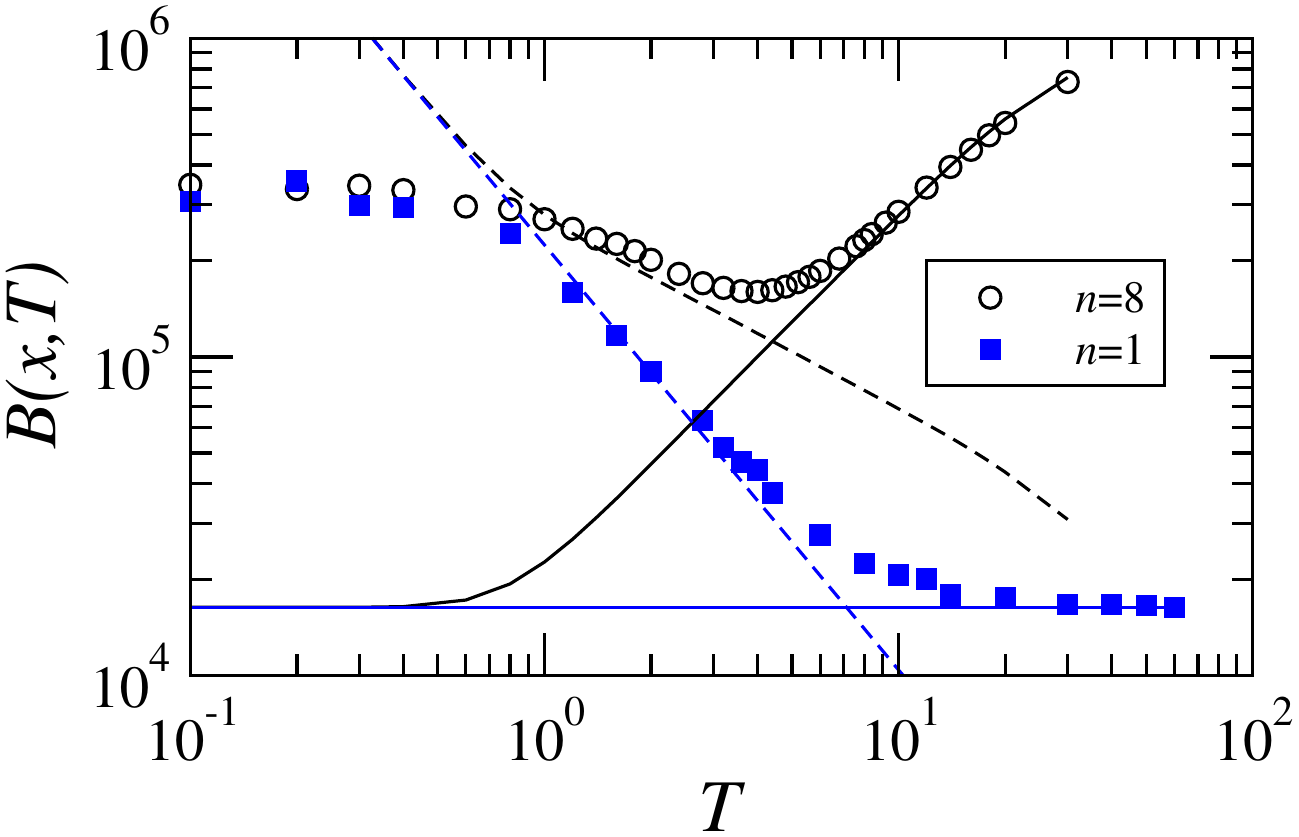}
\caption{\label{fig:BofxT} (Color online) ( (b) Temperature behavior of $B(x,T)$ for fixed $x=16384$. Circles (squares) corresponds to $n=8$ ($n=1$). Continuum lines correspond to the pure thermal regime. Dashed lines correspond to Eq.(\ref{asymp}) with $c=0.54$.}
\label{fig2}
\end{figure}

The weight $Z_{u,x}$ of all polymers starting in $(0,0)$ and ending in $(u,x)$ is given by the following recursion:
\begin{equation}
\label{Zeta}
 Z_{u,x} = e^{-\beta V_{u,x}} \sum_{j=1}^n  e^{-\beta j^2} \left( Z_{u-2 j+1,x-1}+ Z_{u+2 j-1,x-1}\right) \nonumber
\end{equation}
with $Z_{u,0} = \delta_{u,0}$. The probability, for a given disorder realization, to observe a polymer ending in $(u,x)$ is $Z_{u,x}/\sum_u Z_{u,x}$. Since $Z_{u,x}$ grows exponentially we divided all weights at fixed $x$ by the biggest one~\cite{directedpoly} which does not change the ending probability. 
The roughness is defined as the mean square displacement of the free end:
\bea
&& \ B( x) = \overline{\langle  u^2(x) \rangle_{n,T} } =\overline{\sum_u u^2 Z_{u,x}/\sum_u Z_{u,x}}.
\eea
In the rescaled variable the roughness takes the form $\tilde B(\tilde x) = H(\tilde x)$, where $H$ is  universal once the boundary conditions are specified. Here we have  $H(\tilde x \to 0)=\tilde x$ and $H(\tilde x \to \infty)= c \, \tilde x^{4/3}$, with $c$ is a universal constant.  Hence in the original variables one has:
\begin{equation}
 B(x,T) = a^2 H(x/b) = \frac{T^6}{\kappa_n^2 g^2} H\left( x \frac{\kappa_n g^2}{T^5} \right).
\label{collapse}
\end{equation}
This function has a crossover between a pure thermal and a strong disorder regime for $x\gg T^5/\kappa_n g^2$ where:
\begin{equation}
 B(x,T) \to c \,  \left(\frac{g}{\kappa_n T}\right)^{2/3}  x^{4/3}.
 \label{asymp}
 \end{equation}
A decay of the roughness amplitude is thus expected at high temperature. This non-intuitive behavior in $T$ was pointed out in the past \cite{nattermannshapir,us3}, but never observed in numerical simulations, since high enough temperatures where not reached (see Ref.~\cite{iguain2009} and references therein). Here we further check the dependence in $\kappa_n$ and $g$. In Fig.~\ref{fig:BofxT} we show  $B(x,T)$ as a function of the temperature for a {\it fixed} large value of $x$. Three regimes are observed: i) Low temperature $T \le T_{dep}$ (here we estimate $T_{dep} \sim 1$): $B(x,T)$, at this large scale, is temperature independent. ii) Roughness decay: for larger temperature $B(x,T)$ is a decreasing function of $T$ well fitted by the asymptotic behavior of Eq.(\ref{asymp}). iii) Very high temperature $T^5 \gg \kappa_n g^2 x$: the pure thermal regime $B(x,T)=T x / \kappa_n$.  For $n=1$, the roughness achieves a flat plateau, while for $n>1$ as well as for the continuum elastic string  a non-monotonic behavior is expected as shown here for $n=8$ (for any finite $n$ the increasing roughness finally reaches a plateau not shown in the figure). Let us stress that if $x$ is not large enough the roughness decay is not observed and the polymer crosses directly from a low temperature to a pure thermal regime.
 
\begin{figure}[!tbp]
\includegraphics[width=8cm,clip=true]{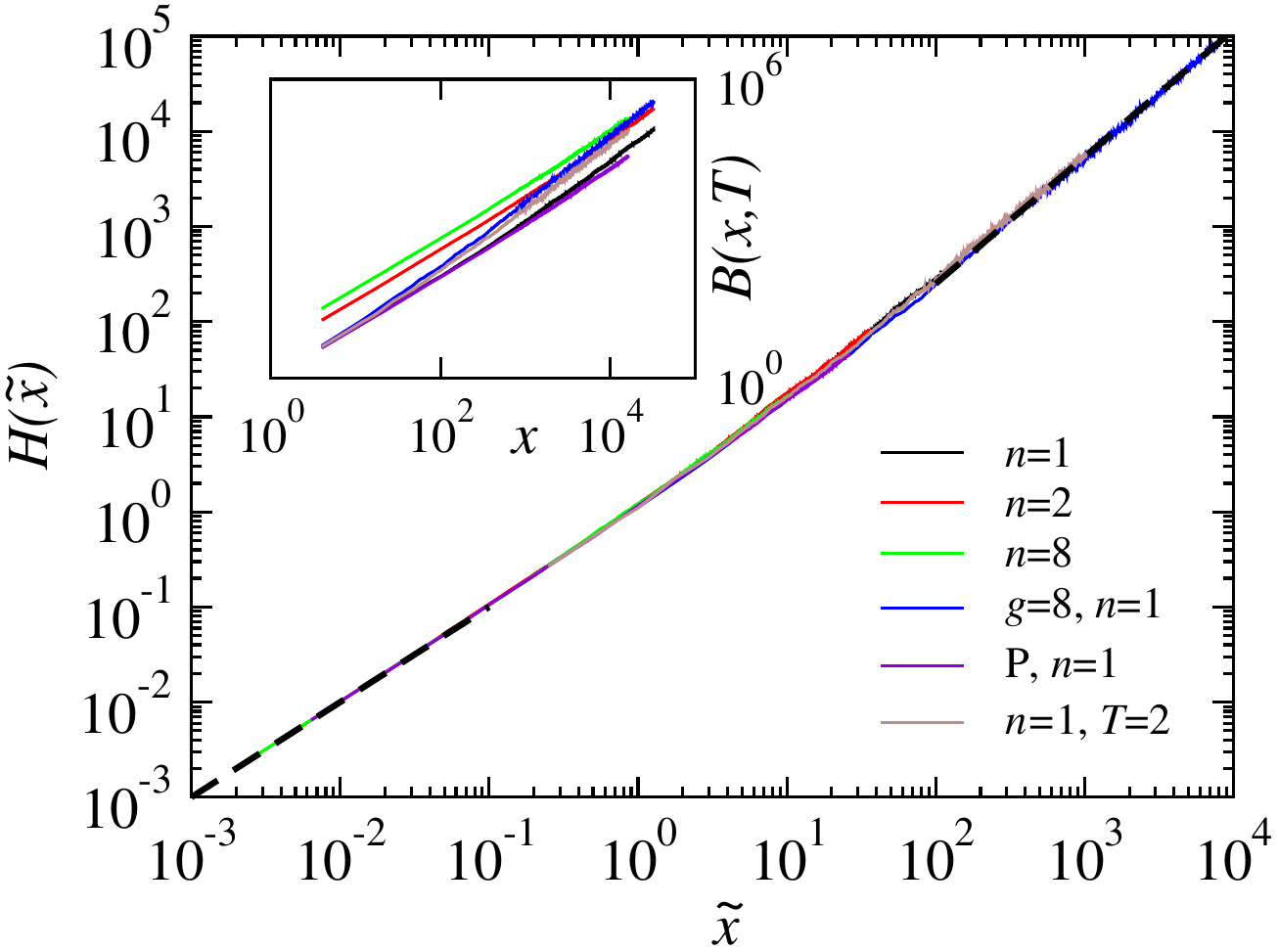}
\caption{\label{fig:s-univ} (Color online) Inset: Raw data for $B(x,T)$. Main: Collapse using Eq. (\ref{collapse}). Dashed lines emphasize the two limit behaviors.  $T=4$ except for the last data set, with $T=2$. P stands for Poissonian disorder ( with $d=2$ and  $p=0.2$).}
\end{figure}

Finally, we test the full scaling relation (\ref{collapse}). The inset of Fig.~\ref{fig:s-univ} shows the raw data for different parameters. As observed in the main panel of Fig.~\ref{fig:s-univ}, using  Eq.\eqref{collapse} and no free parameter, all the data are perfectly collapsed into a single universal function, $H(\tilde x)$, with the fitted value $c = 0.54 \pm 0.01$. This amazing scaling rely on the high temperature properties. For the same models, at $T=0$, there is not a universal function $H_0$ and two model-dependent constants $a$,$b$ such that $B(x)=a^2 H_0(x/b)$.

We now study the high $T$ regime using the FRG. The rescaled disorder correlator $\tilde R(u)$, defined through $R(u) = A_d^{-1} x_0^{d-4} \kappa^{2} e^{- (\epsilon - 4 \zeta) \ell} \tilde R(u e^{- \zeta \ell})$ ($A_d$ a constant), satisfies, under coarse graining over short scales $x < x_0 e^{-\ell}$, the flow equation \cite{BalentsFisher,frg}:
\bea 
&& \! \! \! \! \! \! \partial_\ell \tilde R(u) = (\epsilon - 4 \zeta)  \tilde R(u) + \zeta u_i  \tilde R'_i(u) + \tilde T e^{-\theta l} \tilde R''_{ii}(u) + \beta[\tilde R] \nonumber  \\
&& \! \! \! \! \! \beta_{1loop}[\tilde R] = \frac{1}{2} \tilde R''_{ij}(u) R''_{ij}(u) - \tilde R''_{ij}(u) \tilde R''_{ij}(0)  \label{frg1} 
\eea
where $R'_i\equiv \partial_{u_i} R$, $\tilde T= B_d x_0^{2-d} T/\kappa$ and we indicated the beta function to one loop. The key point is that at high temperature $T$, there is a first regime of the flow, $0 < \ell < \ell_T$, where the non-linear terms are negligible. At $\ell > \ell_T$ they match the rescaling ones, leading to the approach to the $T=0$ fixed point as $\ell \gg \ell_T$. Although noticed in \cite{mueller2001}, all consequences have not been discussed. 

For $\ell < \ell_T$ it is simpler to integrate the flow equation of the unrescaled $R(u)$, $\partial_\ell R(u) = \tilde T e^{(2-d) \ell} R''_{ii}(u)$, hence $$R_\ell(u) = (4 \pi D_\ell)^{-N/2} \int du' e^{- (u-u')^2/4 D_\ell} R_0(u')$$
with $D_\ell = \tilde T \int_0^{\ell} e^{(2-d) \ell}$. We first discuss $d < 2$. Then $D_\ell \approx \tilde T e^{(2-d) \ell}/(2-d)$  grows hence $R_\ell(u) \to g \frac{e^{- u^2/4 D_\ell} }{(4 \pi D_\ell)^{N/2}}$ where here we define $g=\int du' R_0(u')$, which is the only memory of the initial condition. This initial regime involves Flory exponents, i.e. if we choose $\zeta=\zeta_F$, we see that $\tilde g = \int du' \tilde R(u')$ is preserved, and:
\bea
 \tilde R_\ell(u) \to \tilde g \xi_\ell^{-N} f(\frac{u}{\xi_\ell}) \label{solu1}, \; \xi_\ell = \sqrt{D_\ell} e^{- \zeta_F \ell} \sim \sqrt{T} e^{-\frac{\theta_F}{2} \ell} 
\eea
with the "fixed point" shape $f(x) =  (4 \pi)^{-N/2} e^{-x^2/4}$ and $\tilde g=A_d x_0^{4-d} g \kappa^{-2}$. Inserting (\ref{solu1}) into (\ref{frg1}) the rescaling and temperature terms are of the same order, while the $(\tilde R'')^2$ is smaller by a factor $\tilde g/\xi_\ell^{4+N}$~\cite{note1}. The length $\ell_T$ is thus determined as $\tilde g/\xi_\ell^{4+N} \sim 1$. Defining $b  = x_0 e^{\ell_T}$ one recovers formula (\ref{solution}) up to universal constants ($x_0$ cancels). For $\ell > \ell_T$ one can choose the asymptotic value for $\zeta$ and the asymptotic form for the flow is then $R''_\ell(u) = e^{- \epsilon \ell + 2 \zeta (\ell - \ell_T) + 2 \zeta_F \ell_T} \tilde R''(u e^{- \zeta(\ell - \ell_T) - \zeta_F \ell_T})$. This gives  the predicted roughness  decay as:
\bea
 \overline{u(-q) u(q)} \sim \frac{- R_{\ell=\ln(a/q)}''(0)}{q^4}  \to 
 \frac{T^{- 2 \frac{\zeta - \zeta_F}{\theta_F}}}{q^{d+2 \zeta}} 
\tilde R^{* \prime \prime}(0) 
\label{rough}
\eea
Furthermore, the flow enters the low temperature region where non linear terms are important {\it with a $T$-independent initial condition} since the shape of $\tilde R(u)$ has converged to $f(x)$. The  unique RG trajectory  explains the universality in the function $H(\tilde x)$ and others.

For $d > 2$, $D_\ell \to \tilde T/(d-2)$ which is large at high $T$. Universality still holds, with (\ref{solu1}) and $\xi_\ell \approx \frac{1}{d-2} \tilde T e^{- \zeta_F \ell}$. Inserting (\ref{solu1}) into (\ref{frg1}) the temperature term is negligible compared to the rescaling term, but $(R'')^2$ is again exactly down by a factor $\tilde g/\xi_\ell^{4+N}$. This leads to the result for $b$ given in (\ref{solution}) with $\theta_F \to 2 \zeta_F$ , provided that in the initial model (\ref{srep}) one replaces $\kappa/T \to x_0^{2-d} \kappa/T$ and $g/T^2 \to x_0^{2(2-d)}  g/T^2$ to eliminate the $x_0$ dependence in $b$. This implies that for $d>2$, at high $T$, the continuum model  is the small $\tilde x_0$ (rescaled cutoff) limit of:
\be \nonumber
S^{rep} = \frac{\tilde x_0^{2-d}}{2} \int_{x} [ \sum_{\alpha=1}^p \left(\nabla_{x} \tilde u_\alpha(x) \right)^2 -  \tilde x_0^{(2-d)} \sum_{\alpha \beta=1}^p  \delta^N( u_{\alpha \beta})]
\ee
Moreover the roughness decay (\ref{rough}) holds if $\theta_F \to 2 \zeta_F$. 

Let us discuss LR disorder $R(u) \sim 1/u^{\gamma}$ at large $u$. The roughness is known \cite{BalentsFisher} to be given by the LR Flory exponent $\zeta_{LR}=\zeta^F_{LR}=\frac{4-d}{4+\gamma}$ for $\gamma<\gamma_c(N)$ such that $\zeta_{LR} > \zeta_{SR}$. This means that the roughness decay (\ref{rough})  is not present for LR disorder, e.g. random field  disorder. The condition $g$ finite implies $\gamma>N$ hence $\zeta_{LR} < \zeta^F_{SR}$. Since $\zeta^F_{SR} < \zeta_{SR}$, a finite $g$ guarantees that at high $T$ the system is described by the SR fixed point. Similar conclusions hold for LR correlated disorder in $x$ \cite{fedo}. 

We have demonstrated numerically on the DP that a universal high $T$ regime exists, described by a two parameter $\delta$-correlated model. The dimensional and FRG arguments show that this universality within the glass phase, e.g. $B(x)=a^2 H(x/b)$ for the roughness and $\overline{F^2}^c=T^2 F(x/b)$ for the free energy variance, extends to general $d$ ($\theta_F>0$) and boundary conditions.

This work was supported by ANR grant 09-BLAN-0097-01/2 and by MINCYT-ECOS No. A08E03.


\begin{thebibliography}{32}
\bibitem{note1}
Higher loops are also  ordered in powers of this parameter.
\end{thebibliography}
\end{document}